# Balancing Accuracy and Diversity in Recommendations using Matrix Completion Framework

*Anupriya Gogna and Angshul Majumdar*

*Abstract—* **Design of recommender systems aimed at achieving high prediction accuracy is a widely researched area. However, several studies have suggested the need for diversified recommendations, with acceptable level of accuracy, to avoid monotony and improve customers' experience. However, increasing diversity comes with an associated reduction in recommendation accuracy; thereby necessitating an optimum tradeoff between the two. In this work, we attempt to achieve accuracy-diversity balance, by exploiting available ratings and item metadata, through a single (joint) optimization model built over the matrix completion framework. Most existing works, unlike our formulation, propose a 2-stage model - a heuristic item ranking scheme on top of an existing collaborative filtering technique. Experimental evaluation on a movie recommender system indicates that our model achieves higher diversity for a given drop in accuracy as compared to existing state of the art techniques.**

## 1. INTRODUCTION

Use of recommender systems (RS) in online portals has been on a steady rise owing to their utility in mitigating the problem of information overload. This has powered deep interest in design of efficient RS design algorithms for generating product/service recommendations [1, 2, 3, 4].

The accuracy of rating prediction is one of the most widely used measures for determining the efficiency of recommender systems [3, 5]. Existing works on RS design are largely focused on utilizing the available rating information, given by users in the past, to determine their predilection; de-facto approach for the same is collaborative filtering (CF) [6, 7]. CF techniques use the rating information (explicit or implicit) given by users, in the past, to a subset of items to generate future recommendations. Their key focus is on improving accuracy of recommendation [7, 8, 9, 10], which translates into recommending items very similar to the items preferred (rated highly) by users in the past.

Several studies [11, 12, 13] have suggested that such a restricted evaluation of efficiency of recommendation suffers from several limitations. From a user's perspective, generating recommendations very similar to their choice in the past creates monotony and over time RS loses its value. Ideally, RS should be able to provide users with personalized (item) suggestions that will be difficult for them to search on their own. However, most algorithms are biased towards highly popular or frequently rated items, thereby recommending them to a number of users. Such items have enough visibility themselves and thus the prime purpose of RS is defeated.

Further, from the business viewpoint as well, similarity of recommendations is a holdup in achieving desired visibility for a wide range of items. Higher the number of items suggested, especially those with lower visibility, greater are the chances that they are purchased by the user, increasing profit margin for online portals [14, 15]. Also, suggesting varied yet relevant items maintains users' interest and faith in the RS, which has direct impact on profitability of e-portals.

In the light of shortcomings highlighted above, there is a growing impetus towards design of more comprehensive recommender systems, which give sufficient importance to improving the breadth and diversity of recommendation. However, this increase comes at the cost of reduced accuracy and thus focus of current research works [15, 16, 17, 18] is on maximizing the variety in recommendations for an (given) acceptable loss in prediction accuracy.

Existing works on balancing accuracy and diversity of recommendation can be divided into two categories – two stage recommendation strategies and unified models. The former class of work [16, 22, 23] build a cascaded system wherein the first stage uses an existing CF technique to predict the missing ratings and the second level consists of a modified ranking strategy (unlike conventional ranking in order of decreasing rating value) which promotes desired diversification. The second step requires selection of heuristic ranking threshold; items rated above the threshold are considered as prospective candidates for recommendation. Although, the two stage process has the advantage/freedom of using an off the shelf CF technique, it cannot guarantee optimality of solution and also leads to increase in expense. On the other hand, unified models [20, 24] present a joint, single-stage framework consisting of a weighted combination of diversity and accuracy promoting functions. These models primarily rely on using/developing a mathematically sound optimization strategy for solving the joint formulation and therefore, provide some guarantee on the recovery of optimal solution. Despite the theoretically sound framework presented by these models there has been limited exploration of such architectures. Our proposed work belongs to the class of unified models.

Most existing works evaluate the capability of a RS to diversify recommendations on two criteria - individual diversity (ID) [19] and aggregate diversity (AD) [16]. Individual diversity [20] refers to the breadth or range of (varied) items suggested to a given user. It is measured as average pairwise dissimilarity amongst the items recommended to a user. On the other hand, aggregate diversity [21] is a measure of diversity/breadth of recommendations across users and is more attuned to viewing the problem from a system's viewpoint.

Although both ID and AD are diversity centric measures, an increase in one does not guarantee a high value for the other. For example, same set of diverse items (say one from each genre for movie recommendations) could be suggested to all users which will heighten individual diversity but it will not be profitable from the system's perspective. Thus, to ascertain profitability for online portals in addition to improving customer satisfaction, a joint focus towards improving both ID and AD is imperative.

In this work, we present a single stage strategy to jointly optimize accuracy and diversity (individual as well as aggregate). We propose an (convex) optimization framework which establishes a balance between the two desired (yet conflicting) characteristics – accuracy and diversity - while generating rating prediction. In addition to exploiting the available rating values, we also use item metadata (in our case movie genres) to explicitly enforce desired diversifying criteria. To the best of our knowledge, there are very few works in the past [19, 36, 38] which exploit available secondary data (metadata) to establish diversity-accuracy tradeoff. In [19], authors use user tagging (user metadata) information to develop expansion strategy for user interest and suggest diversified items. In [36], taxonomic description of books from amazon is exploited to introduce diversity in recommendation for the book crossing dataset via topic diversification. Similar to our framework, in [38] genre information of movies from Movielens database is used to introduce diversity and ensure that recommendations span multiple genres. These methods, albeit based on introducing diversity quantified in terms of desired criteria (derived from metadata), lacks the robustness and effectiveness of pure optimization based scheme offered by our formulation. These works are discussed in further detail in section 2. Amalgamation of diversity promoting criteria, in the proposed optimization framework (cost function) during the process of rating prediction itself eliminates the need for arbitrary selection of ranking thresholds and heuristic ranking strategies. The ratings predicted by our model are used to rank items in the conventional manner - highest rated is ranked 1st. The convexity inherent in our proposed formulation ensures optimality of solution; thereby yielding desired level of diversity with minimal loss of accuracy – a claim supported by our experimental results. In addition to this, our design yields not only higher individual diversity but higher aggregate diversity and increased novelty of recommendations as well. This is unlike existing models where there is a prominent focus on only one of the diversity measures.

Our model is based on the notion that ideally, to provide maximum diversity, predicted values across diverse item sets should be uniformly distributed [20]. This ensures equal probability of selection from each of the distinct item subsets, thereby enhancing diversity. However, on its own, such random selection can lead to considerable reduction in prediction accuracy. To ensure that accuracy drop is minimized for a given diversity level we formulate an optimization problem combining the uniform distribution constraint with the principles of latent factor model (LFM) based CF design methods.

According to LFM [25], a user's choice of an item is governed by a handful of (defining) factors; such as genre, cast, and language for the case of movie recommendations. The most popular technique, based on LFM, for determining missing ratings is matrix factorization (MF) [26], wherein the rating matrix is recovered as a product of user and item latent factor matrices. A user's latent factor representation captures his/her affinity to these latent factors and that of an item indicates the degree to which various latent factors are possessed by it.

However, MF being bilinear is a non-convex formulation. Recently researchers suggested low rank matrix completion (LRMC) [27, 28] as a convex counterpart of matrix factorization. According to LRMC, missing values in a matrix can be satisfactorily recovered, by looking for minimum rank solution, provided the matrix is adequately low rank. In case of recommender systems, as the entire rating matrix is believed to be governed by a handful of features (latent factors), which do not exceed 50-200 for most cases; the rating matrix (with dimensions running into hundreds of thousand) has (relatively) very low rank. Thus, LRMC techniques can be employed for prediction of missing ratings.

In this work, we build up on the LRMC model (targeted towards higher prediction accuracy) and impose additional constraint which ensures greater diversity in recommendations. This add-on constraint is incorporated into the LRMC framework as a regularization term promoting uniform distribution amongst average ratings across genres. Uniform distribution allows all (varied) genres to have equal chance of representation in the recommendation list; thereby enhancing diversity. Thus, our framework enables simultaneous (joint) optimization of both accuracy and diversity. Although, we have focused on showcasing genre based diversity in movie recommendations in this work, our framework can accommodate any other diversifying criteria derived from various secondary sources.

The novelty of our design lies in proposing a unified convex formulation, utilizing metadata along with rating information, for simultaneously optimizing both accuracy and diversity. We also propose an efficient algorithm for solving our formulation based on split Bregman technique [29]. Use of split Bregman technique ensures improved accuracy and lower computational complexity. There is no existing work in our knowledge that proposes such a framework with an associated algorithm.

The main contributions of our work can be summarized as follows

- Propose a unified framework (single stage strategy) for jointly optimizing accuracy as well as recommendation diversity in RS.
- Propose a convex formulation utilizing item metadata (genre information for movie database) for accuracy-diversity trade-off.
- Design an efficient algorithm using split Bregman technique for our formulation.

The rest of the paper is organized as follows. In section 2, we present a review of existing CF techniques and diversity enhancing measures. Section 3 contains our proposed formulation and algorithm design. In section 4 the experimental set up and evaluation metrics are discussed followed by results. The paper ends with a conclusion in section 5.

## 2.  RELATED WORK

In this section we review the latent factor model and MF approach followed by few existing algorithms for LRMC. We also discuss the prior art aimed at increasing diversity in recommendations.

### 2.1. Matrix Factorization Framework for Latent Factor Model

Latent factor models [25] predict the missing ratings based on the belief that the interaction between users and items is governed by a few features only (numbering ~100 in most cases). Consider as an example movie database such as IMDB. A user's choice for a movie is based on criteria such as cast, genre, director and few others. Similarly, any movie will possess these traits to varying degree. It is not possible to hand-craft the features that guides the user's choice on items; hence latent factor model assumes that the factors are hidden / latent. Every has a certain quantifiable affinity to the factors and every item possesses these factors in a quantifiable amount. Hence, as per the latent factor model, both users and items can be represented by constant length vectors of latent factors [30].

Say $u_i$ denotes the user $i$'s latent factors and $v_j$ the item's. The user rates the item high only if there is a match between $u_i$ and $v_j$; mathematically this is best captured by the inner-product. The rating of $i$ on $j$ is expressed as an inner product between the corresponding latent factors.

$$r_{i,j} = u_i^T v_j \qquad (1)$$

If we consider all users and all items, the entire ratings matrix ($X$) can be expressed as,

$$X = UV \qquad (2)$$

where $U$ is the latent factor matrix of all users and $V$ for all items.

The ratings matrix is only partially observed ($R$); users typically rate only a few movies. Hence the data available to us is represented as

$$R = M(X) = M(UV) \qquad (3)$$

Here $M$ is a binary matrix having 1's where the user rated the item and 0 elsewhere.

Matrix factorization (MF) [26] is the most popular albeit non-convex approach used to compute the user and item latent factor matrices.

$$\min_{U,V} \left\| R - M\left(UV\right) \right\|_F^2 + \lambda \left( \left\| U \right\|_F^2 + \left\| V \right\|_F^2 \right)$$

$where$, $Frobenius\ norm\ of\ a\ matrix\ A \left( \left\| A \right\|_F \right) is\ defined\ s.t.$ \qquad (4)

$$\left\| A \right\|_F^2 \triangleq \qquad {}^T A) = \sum_{i,j} \left| a_{ij} \right|^2$$

Here $\lambda$ is the regularization parameter to prevent over-fitting.

Matrix factorization model (4) is a bilinear, and hence non-convex framework. Although the MF model is a fast and scalable implementation of latent factor framework, the inherent non-convexity does not allow it to enjoy convergence guarantees.

### 2.2. Low Rank Matrix Completion

Usually in collaborative filtering the number of factors are much smaller than the number of users or items. For a typical eCommerce portal the number of users runs into millions and the number of products are of the same order. However the number of factors that decides the users choice on the item are few. For example for book recommendations one typically only considers the genre and author; for movies it is typically the star cast and director. Hence, one can always assume that the number of latent factors are indeed far smaller than the ambient dimensionality of the ratings matrix. In other words the ratings matrix is of low-rank.

Over the years matrix factorization has been a popular technique for low-rank matrix completion. However, in the recent past, researchers have proposed another approach for recovery of missing data in low rank matrices. This is based on searching for a lowest rank matrix among all possible matrices subject to the data fidelity constraint. Ideally one would like to solve,

$$\min_X rank(X) \text{ subject to } R = M(X) \tag{5}$$

Unfortunately this is an NP hard problem with doubly exponential complexity. Theoretical studies in low rank matrix completion (LRMC) [give reference from Candes & Tao / Recht] prove that it is possible to relax the NP hard rank minimization problem by its closest convex surrogate – the nuclear norm.

$$\min_X \|X\|_* \text{ subject to } R = M(X) \tag{6}$$

The nuclear norm $\|\bullet\|_*$ is defined as the sum of singular values. This (6) is a convex problem and can be solved using semi-definite programming.

Usually (6) is not employed. There is noise in the observations, hence the equality constraint is relaxed and the following unconstrained problem is solved instead.

$$\min_X \|Y - M(X)\|_F^2 + \lambda \|X\|_*$$

*where, Nuclear norm of a matrix* $A \left( \|A\|_* \right)$ *is defined s.t.* $\tag{7}$

$$\|A\|_* \triangleq \qquad \overline{A^T A} \Big) = \sum_i \sigma_i(A); \sigma_i(A) \text{ denoting singular values of } A$$

   Nuclear norm minimization requires singular value decomposition at each iteration, which owing to its high computational complexity makes LRMC framework slower compared to MF model. However, LRMC (2) is a convex formulation and thus enjoys proven convergence guarantees, thereby ensuring optimality of solution. Thus, using LRMC assists in yielding more accurate RS design compared to MF formulation.

   Broadly speaking there are two existing approaches to address the problem of LRMC. The first (popular) approach is based on thresholding the singular values [give ref] while the second approach is the iterative reweighted least squares technique [give ref].

   Even though most studies in collaborative filtering are based on the MF framework owing to the ease of solution and faster convergence, we use the LRMC framework for its theoretical advantages and higher accuracies. In this work we speed up LRMC using the split Bregman technique [29] .

## 2.3. Diversity-Accuracy Tradeoff in Recommender Systems

Till recently the prime focus of RS design algorithms was on minimizing error metrics such as RMSE (Root mean squared error) and precision. However, studies like [11] suggested evaluating RS based on other measures such as diversity and unexpectedness in addition to accuracy.

Recommendations made based on accuracy as the only measure will suggest items that are in tune with the user's past history. This may be monotonous for the users after a while. Also from a philosophical standpoint, we expect human beings to be well rounded – we should have a certain curiosity and interest to all possibilities. Wearing the same type of clothes, reading similar books or watching the same genre of movies deter our development. Hence, the argument in [11] to introduce other diversity measures for RS evaluation.

Several attempts have been made to diversify the recommendation list while maintaining adequate relevance to the user's preference. Works such as [16, 21, 33, 34] have focused on the business side alone and aimed to increase the aggregate diversity and/or suggest long tail items. This is another reason (apart from user's well being) to improve diversity from the business perspective. To acquire any item (however obscured), the portal has an associated cost. They need to recover this cost. Hence they need users to buy or rent it. Thus they need to recommend relatively unpopular / obscure items to the users.

Authors in [16] studied the variation in recommendation and item ranking with parameters such as item's net rating, its popularity etc. to come up with several ranking strategies. They proposed a new ranking measure (8)

$$rank\left(i,T_R\right) = \begin{cases} rank_x\left(i\right) & \text{if } R\left(u,i\right) \in \left[T_R, T_{\max}\right] \\ \alpha + rank\left(i\right) & \text{otherwise} \end{cases} \tag{8}$$

where, $T_R$ is the ranking threshold and $rank_x$ is the new ranking strategy applied to items ranked above a specific (chosen) ranking threshold; $rank$ denotes the usual ranking strategy (highest rated ranked first). The heuristic selection of parameters and strategies in the work makes it less effective than pure optimization based schemes. In [21] authors suggested to recommend users to items rather than traditional approach of recommending items to users in order to give fair opportunity to all items. They also designed a probabilistic approach to address the same. Authors in [33] borrowed concepts from association mining to create characterization vectors for long tail items and establish their correlation with more heavily rated items. In [35] the problem of liquidating long tail stock was addressed using relevance models. Authors in [34] used evolutionary algorithm to improve visibility of long tail items, however such a model cannot provide any guarantees on optimality of solution. Also, it is difficult to vary the model parameters to introduce varying degree of diversity. The explicit focus of above works is only on suggesting a larger number of items which may not result in improved diversity for users, thereby affecting their interest in RS.

Other works [36, 37, 38] have addressed the problem from user's perspective and suggested schemes to improve individual diversity. Authors in [36] proposed a topic diversification based method to increase individual diversity. They used hierarchical classification of items combined with item based CF to generate diverse yet reasonably accurate suggestions. The classification is done by using the item background information obtained from its taxonomic description. In [37] the idea of k-furthest neighbors was introduced. They showed that suggesting items disliked by furthest neighbors lead to higher diversity with small reduction in accuracy. Both these methods are based on heuristic measures and neighborhood based techniques which are not as effective as latent factor models. Only a few works [20, 38] have focused on pure optimization based designs. In [38] a probabilistic framework is proposed that aims at generating recommendation list such that the probability that user will select item from the list is maximized. The items (movies in this case) are segregated into multiple lists based on the available genre information. However, unlike our work, they proposed a greedy scheme for solving their optimization framework. Also, their model is suited for binarized ratings alone and suggests only one items at a time to the user. Authors in [20] presented a more cohesive model for accuracy-diversity tradeoff. They formulated a binary optimization problem combining two optimizing measures; one promoting diversity and other accuracy (9).

$$\max_{y} \left(1-\theta\right)\alpha y^T D y + \theta \beta m^T y \quad s.t \quad 1^T y = p \tag{9}$$

where, $y$ is an indicator vector having value 1 at $i^{th}$ position if item $i$ is recommended, $D$ is a matrix defining pair wise distance (dissimilarity) between items and $m$ is the vector defining a user's affinity to each item. They also used a greedy scheme for solving the same owing to quadratic nature of the problem.

In this work we present a single-stage convex optimization framework for simultaneous accuracy-diversity optimization. Convex nature of our framework provides it with proven convergence guarantees. Also, no random selection of parameters is required. Our model leads to substantial increase in both individual as well as aggregate diversity with minimal loss in accuracy. We also exploit readily available item metadata which enables us to explicitly enforce desired diversification criteria; in this work movie genre information is used to introduce genre centric diversity in recommendations. To the best of our knowledge, such a framework, exploiting metadata, has not been proposed before. Although, authors in [19] have used item tagging to expand user interest, however their focus is on individual diversity alone and it's a two stage model unlike our unified approach.

## 3. PROPOSED FORMULATION AND ALGORITHM DESIGN

In this section, we discuss our proposed model to achieve diversity-accuracy balance in recommender systems. The novelty of our work lies in formulating a convex framework which aims to simultaneously optimize accuracy and diversity, using rating information as well as available (item) metadata.

Our formulation belongs to class of latent factor models which is shown to yield higher prediction accuracy than neighborhood based models [4]. However, unlike traditionally employed MF framework for LFMs, we use a low rank matrix completion framework to define the prediction model. The basic LRMC framework is augmented with diversity promoting constraint. In this work, we showcase the performance of our model on movie recommendation; focusing on genre centric diversity i.e. aiming at recommending movies across distinct genres. The diversity constraint is derived from the insight that for maximizing diversity, the ratings across diverse item sets (genre in case of movie recommendation) should be uniformly distributed. As highlighted in further sections, although the results are shown for specific diversity measures, our formulation is a generalized framework for incorporating any diversifying criteria across multiple domains. We also design an efficient algorithm based on split Bregman technique for our formulation.

### 3.1. Proposed Formulation

A user $u$'s explicit rating $\left( R_{u,i} \right)$ for an item $i$ is a result of two factors – baseline estimate and interaction component [26]. Baseline (7) defines the general rating behavior/pattern of an item and/or by a user.

$$Baseline_{u,i} = b_u + b_i + mean_g \qquad (10)$$

In (10), $b_u$ is the user bias; higher (positive) value indicating a user's tendency to give ratings higher than the average while a negative value indicates a critical user, $b_i$ is the item bias which is high for very popular items (as they tend to be rated highly by almost all users) and $mean_g$ is the global mean of the available rating dataset.

Interaction component captures the correlation/match between users and items. WE have discussed it in section 2, but we repeat it for the sake of reader's convenience. According to latent factor model, a user's choice for an item is a function of only a few parameters (e.g. Genre, director, actors for movies; usually ~50). Under this belief, a user can be modeled as a vector $\left( U_u \text{ for user } u \right)$ whose value quantifies a user's like or dislike towards each of the parameters or latent factors. On similar lines, each item can be modeled as a vector $\left( V_i \text{ for item } i \right)$ indicating the degree to which it possesses each of the latent factors. Interaction component of the rating $\left( R_{u,i} \right)$ is defined as

$$Interaction_{u,i} = \left( U_u \right)^T V_i \tag{8}$$

Baseline estimation can be easily done offline using stochastic gradient descent (SGD) algorithm by solving (9).

$$\min_{b_u, b_i} \sum_{u,i \in \Theta} \left( R_{u,i} - b_u - b_i - mean_g \right)^2 + \delta \left( \left\| b_u \right\|_2^2 + \left\| b_i \right\|_2^2 \right) \tag{9}$$

Once (offline) baseline estimation is done, the net interaction component $\left( Y_{u,i} \right)$, of available ratings, can be computed as $Y_{u,i} = R_{u,i} - b_u - b_i - mean_g$. We follow similar strategy in our model and work with interaction component alone. Once the complete interaction matrix is recovered, baseline estimates are added back to generate net predicted rating value. In subsequent section, the word 'rating' and 'interaction component' are used interchangeably.

Conventionally, the complete interaction matrix (Z) is recovered as a product of two matrices (U and V consisting of user's and item's latent factor vectors respectively) using MF approach as follows

$$\min_{U,V} \left\| Y - M \left( UV \right) \right\|_F^2 + \lambda \left( \left\| U \right\|_F^2 + \left\| V \right\|_F^2 \right) \tag{10}$$

Albeit popular, (10) is a bi-linear non-convex formulation with no proven convergence guarantees. In this work, we aim to recover the interaction matrix (Z=UV) directly using low rank matrix completion (LRMC) as in (11).

$$\min_Z \left\| Y - M(Z) \right\|_F^2 + \lambda_n \left\| Z \right\|_* \tag{11}$$

where, $M$ is a binary mask or subsampling operator $\left( M_{u,i} = 1 \text{ if user } u \text{ rated item } i \right)$; $\lambda_n$ is the regularization parameter. The rating matrix is influenced by only a small number of latent factors; compared to the dimensions of the rating matrix (several thousands of users and items), the number of the latent factors or independent variables is very low (~100). Thus, it has a (substantially) low rank structure justifying the use of LRMC framework to recover the interaction matrix.

Equation (11) ensures recovery of interaction matrix with high precision (as indicated by results shown in section 4). However, it places no emphasis on obtaining a diverse set of recommendations.

Our goal is to modify the formulation in (11) to predict ratings such that the top-N recommended items, have representation from diverse item sets while showing adequate alignment to the users' past preferences. In order to obtain maximally diversified recommendations, for a given drop in accuracy, each diverse set of items must have equal probability of being recommended. In this work, we focus on the movie database wherein genre defines the diversification criteria i.e. recommendation list should ideally have equal probability of representation/selection from each genre. It should be noted that genre is relevant for movies (may be for books as well). But our framework is generic and can incorporate other factors as well depending on the problem. For example in garment recommendation, one might want to consider different styles or different colors to improve diversity.

When combined with the base model in (11), this diversifying criterion ensures that diversity is incorporated in recommendations while maintaining adequate relevance to a user's preference (latter credited to the base accuracy centric model).

To accommodate the diversification criteria, we include an additional regularization term in (11) to promote minimum variability amongst average rating (by each user) for each genre as follows

$$\min_Z \left\| Y - M(Z) \right\|_F^2 + \lambda_n \left\| Z \right\|_* + \lambda_d \sum_{u \in users} \text{var}_{genre}(u) \tag{12}$$

where, $\text{var}_{genre}(u) = \sum_{g \in genre} \left( Z^A_{u,g} - m_{u,genre} \right)^2$ denotes the variance of the vector consisting of average ratings for each genre for a given user $u$; $\lambda_d$ is the regularization parameter. $Z^A_{u,g}$ is the average rating by user $u$ for movies belonging to genre $g$ and $m_{u,genre} = \dfrac{1}{|genre|} \sum_{g \in genre} Z^A_{u,g}$ is the mean of ratings across all genres.

The nuclear norm constraint in (12) ensures low rank structure yielding high correlation with previously recorded rating pattern of a user. In contrast, the diversifying criteria - uniform distribution term - in (12) promotes deviation away from set pattern by virtue of spreading the ratings uniformly across genres. The values of $\lambda_d$ and $\lambda_n$ determine the relative importance placed on prediction accuracy (promoted by nuclear norm constraint) and diversity (promoted by variance minimization constraint) in the recommendation list. The optimum value of regularization parameters is determined using $l$-curve technique [39]. As the model is based on a convex optimization framework, optimality of solution i.e. minimum loss in accuracy for a given degree of diversity is ensured.

The variance based regularization term in (12) can be represented in matrix form as in (13) where $m$ is the number of users; $n$ is the number of items; $d = |genre|$ is the number of genre considered; $\mu_g$ is the number of movies belonging to genre $g$ and $\mathbf{1}_{d \times d}$ is a matrix of dimension $d \times d$ consisting of all 1's. Matrix $G$ is defined such that $G_{i,j} = 1$ *iff* movie $i$ belongs to genre classification $j$.

$$\left\| \begin{bmatrix} Z_{11} & Z_{12} & . & Z_{1n} \\ Z_{21} & Z_{22} & . & . \\ . & . & . & . \\ Z_{m1} & . & . & Z_{mn} \end{bmatrix} \underbrace{\begin{bmatrix} \dfrac{G_{11}}{\mu_1} & \dfrac{G_{12}}{\mu_2} & . & \dfrac{G_{1d}}{\mu_d} \\ \dfrac{G_{21}}{\mu_1} & \dfrac{G_{22}}{\mu_2} & & \\ . & & . & \\ \dfrac{G_{n1}}{} & & . & \dfrac{G_{nd}}{} \end{bmatrix}}_{n_{u,g} \vee users,\ g=genre} - \right.$$

$$\left. \dfrac{1}{d} \begin{bmatrix} Z_{11} & Z_{12} & . & Z_{1n} \\ Z_{21} & Z_{22} & . & . \\ . & . & & . \\ Z_{m1} & . & & Z_{mn} \end{bmatrix} \underbrace{\begin{bmatrix} \dfrac{G_{11}}{\mu_1} & \dfrac{G_{12}}{\mu_2} & . & \dfrac{G_{1d}}{\mu_d} \\ \dfrac{G_{21}}{\mu_1} & \dfrac{G_{22}}{\mu_2} & & \\ . & & . & \\ G_{n1} & & . & G_{nd} \end{bmatrix}}_{m_{u,genre} \vee users} \begin{bmatrix} \mathbf{1}_{d \times d} \end{bmatrix} \right\|_F^2 \tag{13}$$

Using (13) in (12), we can formulate our problem as follows

$$\min_Z \left\| Y - M\left(Z\right) \right\|_F^2 + \lambda_n \left\| Z \right\|_* + \lambda_d \left\| Z G_\mu - Z G_\mu \left[ \overline{\mathbf{1}}_{d \times d} \right] \right\|_F^2 \tag{14}$$

where, $G_\mu\left(i,j\right) = \dfrac{G_{i,j}}{\mu_j}$ and $\left[ \overline{\mathbf{1}}_{d \times d} \right]$ is a matrix of dimension $d \times d$ with each element as $\dfrac{1}{d}$.

Equation (14) can be written concisely as

$$\min_Z \left\| Y - M\left(Z\right) \right\|_F^2 + \lambda_n \left\| Z \right\|_* + \lambda_d \left\| ZF \right\|_F^2 \tag{15}$$

where, $F = G_\mu \left( I_{d \times d} - \left[ \overline{\mathbf{1}}_{d \times d} \right] \right)$; $I_{d \times d}$: Identity matrix

Equation (15) represents our proposed (convex) formulation. The resultant matrix $Z$ obtained on optimizing (15) contains ratings (interaction component) recovered such that both prediction accuracy and diversity are jointly optimized.

Once $Z$ is recovered the baseline estimates are added back to determine the (net) predicted rating values. As the predicted ratings are themselves an outcome of balancing diversity and accuracy, there is no need for random/heuristic ranking strategy to be adopted. The ratings are ranked in order of decreasing predicted value to generate top-N recommendation list for each user.

The diversity promoting term in our formulation yields high individual diversity. In addition, as for each individual user, accuracy (recommending preferred genre) is also considered along with diversity (reasonably similar chance of recommendation for each genre), our model is able to achieve high aggregate diversity as well. Consider for example a user who likes comedy, he/she will be recommended more comedy oriented movies but other genres like action, drama will also find representation. Similarly, if another user likes drama, more emphasis will be on movies belonging to this genre. Considering the overall scenario, across all users, many dissimilar movies spanning multiple genres are recommended thereby improving sales diversity. Our claim is supported by the results shown in the following sections.

As mentioned briefly before, our proposed framework can be modified to accommodate other diversifying criteria (beyond genre) across multiple application domains (books, movies, travel plans etc.). For example, in case of book recommendations, we can include diversity in terms of authors, or author demographic./nationalities, or period of the text, or category such as fiction, drama, thriller etc. In such a case, say for suggesting books belonging to various (time) periods, the variance term in (12) can be modified as $\text{var}_{period}(u) = \sum_{p \in period} \left( Z^A_{u,p} - m_{u,period} \right)^2$.

Here various distinct time periods, early 1900 or late 2000, can be considered as diverse subsets and books can be categorized as being written in these time periods. In such a scenario, a user will be suggested books he/she will like across ages, instead of focus on just new releases which anyway have much greater visibility than old texts. In addition, we may club multiple diversifying criteria also to cluster items – say a combination of genre and tagging information for movies. In this case, we may modify our formulation to promote equal representation from each of these clusters. Here the variance term can be modified to minimize variability in average ratings given to items in each cluster. Thus, even though the results showcased are for genre centric diversity, as it is popular and commonly used diversifying criteria [38], our framework is actually a generalized model.

In the next section, we present the design of an efficient algorithm to support our formulation.

*3.2. Algorithm Design*

We design an efficient algorithm for our proposed framework using split Bregman technique [29]. Split Bregman algorithm is an adaptation of alternating direction method of multipliers (ADMM) [40] particularly suited to solving $l_1$ minimization problem. As nuclear norm minimization is an extension of sparse (vector) recovery, split Bregman technique can be well adapted to solving problem with low rank constraints.

Idea behind split Bregman technique is to enable separation of multiple norm terms so that each can be efficiently (and separately) solved. Also, unlike conventional approaches (like iterative soft thresholding [41]) the regularization parameters need not be cooled and can be maintained at optimum value for better convergence and accuracy of recovery. We will not review split Bregman algorithm in detail here, for further details readers may refer [29].

To enable split Bregman type separation of norm terms, we introduce a proxy (to original optimizing variable $Z$) variable $\left(W\right)$ in our formulation as in (!6).

$$\min_{Z,W} \left\|Y - M\left(Z\right)\right\|_F^2 + \lambda_n \left\|Z\right\|_* + \lambda_d \left\|WF\right\|_F^2 + \eta \left\|W - Z - B\right\|_F^2 \tag{16}$$

$$
\begin{aligned}
&\textit{Initialize variables}\\
&\textit{Set parameters}: max\_iter, \lambda_n, \lambda_d, \eta\\
&\textit{while } k < max\_iter \textit{ or } obj\_func(k) - obj\_func(k-1) \le 1e\text{-}7\\
&//\textit{ Solve for } Z\\
&Z \leftarrow Soft\left( sing\_val\left[ Z + \frac{1}{\alpha}\left( \begin{pmatrix} M \\ \sqrt{\eta}I \end{pmatrix}^T \left( \begin{pmatrix} Y \\ \sqrt{\eta}\,(W-B) \end{pmatrix} - \begin{pmatrix} M \\ \sqrt{\eta}I \end{pmatrix} Z \right) \right) \right], \frac{\lambda_n}{2\alpha} \right)\\
&//\textit{ Solve for } W\\
&\textit{Solve } W\left( \eta I + \lambda_d FF^T \right) = \eta\left( Z + B \right)\\
&//\textit{ Update Bregman variable}\\
&B = B + Z - W\\
&\textit{end while}
\end{aligned}
$$

Fig 1. Algorithm for Matrix Completion with Accuracy-Diversity Balance (MCAD)

In (16), instead of imposing a strict constraint of $W = Z$, we use augmented lagrangian type formulation where, $B$ is the Bregman variable (used to compensate for difference between $Z$ and its proxy $W$); $\eta$ is the regularization parameter. Use of Bregman variable removes the need to impose equality constraint between $W$ and $Z$ from first iteration itself. As iterations proceed Bregman variable is updated so that error/difference between Z and W is minimized. Updation of Bregman variable internally as part of the algorithm ensures that recovery error is minimized and convergence behavior improves.

Next, as the two variables W and Z are separable, we split the problem in (16) into two simpler sub problems each minimizing over a single variable as follows.

Sub problem 1:

$$
\min_Z \left\| Y - M(Z) \right\|_F^2 + \lambda_n \left\| Z \right\|_* + \eta \left\| W - Z - B \right\|_F^2 \tag{17}
$$

Sub problem 2:

$$
\min_W \lambda_d \left\| WF \right\|_F^2 + \eta \left\| W - Z - B \right\|_F^2 \tag{18}
$$

Considering sub problem 1 (17), it can be recast as

$$\min_{Z} \left\| \begin{pmatrix} Y \\ \sqrt{\eta}\,(W-B) \end{pmatrix} - \begin{pmatrix} M \\ \sqrt{\eta}\,I \end{pmatrix} Z \right\|_{F}^{2} + \lambda_{n} \|Z\|_{*} \tag{19}$$

Equation (19) is a nuclear norm minimization problem which can be solved using soft thresholding of singular values [42] as in (20).

$$soft \left( singular\ value \left[ Z + \frac{1}{\alpha} \begin{pmatrix} \begin{pmatrix} M \\ \sqrt{\eta}\,I \end{pmatrix}^{T} \\ \left( \begin{pmatrix} Y \\ \sqrt{\eta}\,(W-B) \end{pmatrix} - \begin{pmatrix} M \\ \sqrt{\eta}\,I \end{pmatrix} Z \right) \end{pmatrix} \right], \frac{\lambda_{n}}{2\alpha} \right) \tag{20}$$

where, $Soft(t,u) = sign(t) \ \max(0, |t|-u)$ and $\alpha \geq \max\left( eig \begin{pmatrix} M \\ \sqrt{\eta}\,I \end{pmatrix}^{T} \begin{pmatrix} M \\ \sqrt{\eta}\,I \end{pmatrix} \right).$

Equation (18) is a simple linear system of equation (21) which can be solved using any gradient type solver; we used lsqr [43].

$$W\left(\eta I + \lambda_{d} F F^{T}\right) = \eta\left(Z + B\right) \tag{21}$$

The two sub problems, (20) and (21) are alternately solved with consecutive iteration of both interlaced with updation of Bregman variable as in (22)

$$B = B + Z - W \tag{22}$$

The iterations continue until convergence, i.e. maxi-mum number of iterations reached or reduction in objective function value drops below the threshold. The complete algorithm is summarized in figure 1.

## 4. EXPERIMENTAL SETUP AND RESULTS

We conducted experiments on the real world movie database from Grouplens - Movielens 100K and 1M datasets (http://grouplens.org/datasets/movielens/). They are the most widely used publically available test set for determining the performance of recommender systems. We used the item metadata (genre information) available in these datasets for our proposed framework. The Movielens database also contains a larger 10M dataset, however lack of any secondary information in the same renders it unusable for our formulation. We compared the performance of our algorithm against several existing works to demonstrate the relevance and superiority of our design in suggesting accurate as well as diverse recommendation list; reinforcing customer's satisfaction.

*4.1. Description of Dataset*

The 100K dataset has ratings by 943 users on 1682 movies, with each user giving at least 20 ratings. The 1M dataset has around 3900 movie ids, 6040 users and 1 million ratings.

Item metadata (genre information) is available for all the movies in both the datasets. There are 18 different genres into which movies are clubbed – Action, Adventure, Animation, Children's, Comedy, Crime, Documentary, Drama, Fantasy, Film-Noir, Horror, Musical, Mystery, Romance, Sci-Fi, Thriller, War and Western.

*4.2. Experimental Setup and Evaluation Metrics*

We conducted fivefold cross validation on both the datasets with 80% data used for training and 20% reserved for testing. For each algorithm, on each test set, 100 simulations were carried out and the results shown are aver-aged values for all runs. The simulations are carried out on system with i7-3770S CPU @3.10GHz with 8GB RAM.

The evaluation of our framework and other algorithms compared against was done on the basis of achieved tradeoff between prediction accuracy and diversity measures.

To evaluate prediction accuracy for top-N recommendation approach two quantitative measures are used; namely precision and recall [44]. Precision (23) measures the probability that a recommended item is relevant to the user. We categorize an item in test data as relevant if it's rated 4 or higher (out of 5) by the user as is common in RS evaluation for Movielens dataset.

$$Precision = \frac{\# t_p}{\# t_p + \# f_p} \tag{23}$$

where, $t_p$ defines true positive i.e. items that are relevant to the user and are selected by RS; $f_p$ defines false positives i.e. irrelevant items recommended to the user. Precision usually drops with increase in length of recommendation list.

Recall (24) measures the chances that a relevant item will be selected for recommendation.

$$Recall = \frac{\# t_p}{\# t_p + \# f_n} \tag{24}$$

where, $t_p$ defines true positive; $f_n$ defines false negatives i.e. relevant items not recommended to the user. The recall value usually increases with increases in recommendation list length.

Further rating prediction accuracy is also measured in terms of Mean absolute error (MAE) (25) and Root mean squared error (RMSE) (26).

$$MAE = \frac{\sum_{i,j} R_{i,j} - \hat{R}_{i,j}}{|R|} \qquad (25)$$

$$RMSE = \sqrt{\frac{\sum_{i,j} \left( R_{i,j} - \hat{R}_{i,j} \right)^2}{|R|}} \qquad (26)$$

where, $R$ and $\hat{R}$ are the actual and predicted ratings and $|R|$ is the cardinality of the rating matrix $R$.

To evaluate the measure of diversity introduced by an algorithm, we compute four measures – individual diversity (to gauge diversity from user's perspective), aggregate diversity (to gauge diversity form business perspective), novelty and Gini coefficient (to measure the impact on long tail items).

Individual diversity (ID) [22] or intra-list similarity is a function of variety/diversity offered to a given user. It is measured in terms of (dis)similarity amongst the items in the recommendation list for a given user; net ID computed as an average of the ID for each user (27).

$$Individual\ Diversity =$$

$$\frac{1}{|Users|} \sum_{u \in Users} \frac{\sum_{i \in RL(u)} \sum_{j \in RL(u)} \left( 1 - sim(i,j) \right)}{N(N-1)} \qquad (27)$$

where, $|Users|$ is the number of users; $RL(u)$ denotes the recommendation list of the $u^{\text{th}}$ user; $N$ is the length of recommendation list; $sim(i,j)$ is the similarity between item $i$ and item $j$ in the recommendation list - we used cosine similarity to compute the same.

Aggregate diversity [16] measures diversity from a business perspective. It quantifies the extent to which different items are recommended. It is measured as the total number of unique items from the entire repository recommended across all users (28).

$$Aggregate\ Diversity = \left| \bigcup_{u \in} \right. \qquad (28)$$

Equation (28) defines cardinality of union of recommendation list of all users.

We also measure the novelty [14] i.e. newness of the items recommended. A higher value of novelty indicates that a higher number of long tail items are recommended. Thus, a more novel RS improves visibility of long tail or niche products which also improves profit margins for the retailers. Also, it serves the purpose of RS by suggesting those items to users which they wouldn't have usually found out on their own. Novelty is defined as

$$Novelty = \frac{1}{|Users|} \sum_{u \in Users} \frac{\sum_{i \in RL(u)} \log_2 \left( \frac{|Users|}{\#i} \right)}{|RL(u)|}$$ (29)

where, $|Users|$ is the number of users; $RL(u)$ denotes the recommendation list of user $u$; $\#i$ is the number of ratings for item $i$ available in the training data, a higher value indicating a more popular or commonly rated item. In (29), the logarithm term, and in effect novelty, increases as higher number of niche products (indicated by lower value of $\#i$ ) are recommended.

The diversity in recommendation is also measured using conventionally adopted Gini coefficient [44] defined as

$$Gini\ Coefficient = \frac{1}{J} \sum_{k=1}^{J} \left( 2k - J - 1 \right) p\left( i_k \right)$$ (30)

where, $J$ is the total number of distinct items and $p\left( i_k \right)$ is the probability of $k^{th}$ item being recommended when arranged in increasing order of probability. To compute the Gini coefficient, probability of each item being recommended is computed as

$$p\left( i \right) = \frac{no.\ of\ times\ item\ i\ recommeded\ across\ all\ users}{|Users|\ N} ;\ |Users|$$ denotes the total number of

users and $N$ is the number of items recommended to each user. The probability values thus obtained are arranged in ascending order with $p\left( i_k \right)$ denoting the probability value at $k^{th}$ location in the ordered list. Smaller is the value of Gini coefficient, more uniform is the probability distribution characterizing the chance of each item being recommended; thus, a lower value of Gini index points towards higher diversity.

In most works, using Movielens dataset for testing, the recommendation list length between 1 and 20 is considered [] to compute various evaluation measures such as precision and recall. Following the convention, we show detailed results for a recommendation list of length 5. To highlight the robustness and effectiveness of our model across varying lengths of recommendation list, we also report a summary of results for a recommendation list of length 10 and 20 in sub-section 4.6.

### 4.3. Efficiency of our base Matrix completion model

In this section, we illustrate the effectiveness of our base model – LRMC framework, repeated here for convenience.

$$\min_{Z} \left\| Y - M(Z) \right\|_F^2 + \lambda_n \left\| Z \right\|_* \qquad (31)$$

Equation (31) is also solved using split Bregman technique, on similar lines as discussed in section 3 for our proposed model. We show that our base formulation – MC (31), focused towards improving recommendation accuracy yields better results than conventionally employed matrix factorization techniques. We compare our MC model against existing state of the art MF approaches – Block co-ordinate descent based non negative matrix factorization (BCD-NMF) [45] and Parametric Probabilistic Matrix Factorization (PPMF) [46] for the same, and factored item similarity model (FISM) [47].

TABLE 1. COMPARISON OF OUR BASE MC MODEL WITH MF ALGORITHMS - 100K DATASET (N=5)

| Algorithm | Precision | Recall | MAE | RMSE |
|-----------|-----------|--------|--------|--------|
| MC | **0.8148** | **0.2284** | **0.7351** | **0.9319** |
| PPMF | 0.8067 | 0.2243 | 0.7441 | 0.9601 |
| BCD-NMF | 0.7877 | 0.2221 | 0.7582 | 0.9816 |
| FISM | 0.7801 | 0.2011 | 0.7432 | 0.9439 |

TABLE 2. COMPARISON OF OUR BASE MC MODEL WITH MF ALGORITHMS - 1M DATASET (N=5)

| Algorithm | Precision | Recall | MAE | RMSE |
|-----------|-----------|--------|--------|--------|
| **MC** | **0.8716** | **0.1918** | **0.6813** | **0.8711** |
| PPMF | 0.8691 | 0.1917 | 0.7082 | 0.9013 |
| BCD-NMF | 0.8684 | 0.1921 | 0.6863 | 0.8790 |
| FISM | 0.8578 | 0.1899 | 0.7196 | 0.9102 |

Table 1 and 2 show the comparison of various algorithms in terms of accuracy centric evaluation metrics for 100K and 1M datasets respectively for recommendation list of length i.e. N=5. It can be observed from the reported values that our base matrix completion approach (MC) gives highest rating prediction accuracy amongst all approaches. The precision and MAE obtained with MC shows an improvement of ~1% over the best MF approach compared against. As our purpose is to maximize diversity while retaining a high value of accuracy, our use of MC approach as base model is justified (as it provides highest precision over which diversity enhancement can be applied).

*4.4 Impact of our proposed Model – MCAD – in Improving Diversity*

In this section, we compare the performance of our proposed model – MCAD (15), in terms of its ability to diversity recommendations with minimal loss of accuracy compared to the base MC formulation (31). Both the formulation, MC and MCAD are solved using split Bregman approach; thereby the comparison is indicative of the model design only. As MC is shown to be the most accurate model (from results in table 1 and 2), we only show comparison with it in this section. All the results shown henceforth are for a recommendation list of length 5 unless specified otherwise.

TABLE 3. COMPARISON OF MC BASELINE WITH MCAD FOR MOVIELENS DATASET (N=5)

| Algorithm | Precision | Recall | Aggregate Diversity | Individual Diversity | Novelty | Gini Coeff. |
|---|---|---|---|---|---|---|
| **For 100K dataset** | | | | | | |
| MC | 0.8148 | 0.2284 | 274 | 0.3152 | 1.3011 | 0.9318 |
| **MCAD** | 0.7800 | 0.2187 | **467** | **0.3609** | **1.6981** | **0.8700** |
| **For 1M dataset** | | | | | | |
| MC | 0.8716 | 0.1918 | 652 | 0.3459 | 1.5584 | 0.9598 |
| **MCAD** | 0.8601 | 0.1905 | **1482** | **0.3943** | **1.9950** | **0.8716** |

Table 3 shows the comparison between MC and MCAD model for 100K and 1M Movielens datasets in terms of various accuracy and diversity centric measures. As the focus of our MCAD model is to generate top-N (items) recommendation list which balances accuracy and diversity we compare accuracy in terms of precision and recall measures only; MAE and RMSE are not used, they being rating accuracy centric criteria.

It can be observed from the given results that MCAD is able to achieve significantly higher diversity (individual as well as aggregate) with a very small loss in precision over the base matrix completion (MC) algorithm. We are able to achieve an increase of 51% in aggregate diversity, 22% in novelty and ~7% reduction in Gini coefficient for a small 3% loss in precision over MC algorithm for 100K dataset. Similar observations can be made for the 1M dataset as well. For just a small reduction of around 1% in precision we are able to increase aggregate diversity by 127%, individual diversity by 13%, novelty by 28% and Gini index reduced by 26%.

Most recommender system algorithms are biased towards heavily rated or popular items thereby recommending them across multiple users. This leads to build up of the long tail items in the repository. Fig. 2 shows the impact of our MCAD model in flattening the distribution of number of times an item is recommended for top 400 and top 800 items for 100K and 1M datasets respectively. It can be observed that our algorithm is able to flatten the distribution; reducing the maximum number of recommendations for any item as well as increasing the number of items recommended.

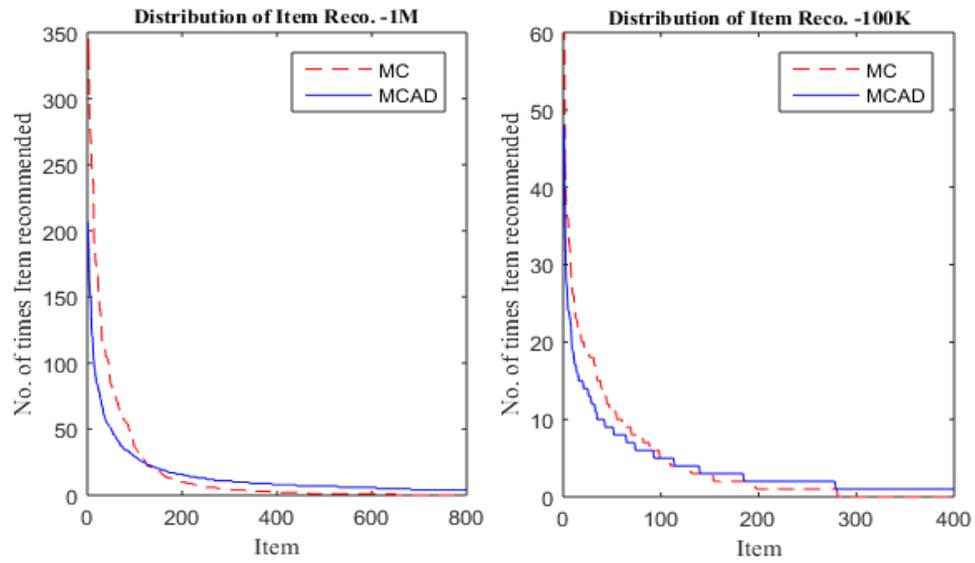

Fig 2. Distribution of number of recommendation for top-T recommended Items

Fig. 3 shows the details of the distribution for 100K dataset. Top figure shows the distribution for top-30 recommended item whereas bottom one shows the same for bottom 30 recommended items using MCAD algorithm. For comparison, the number of times the same item set is recommended using MC algorithms is also shown.

It can be seen from above figure that our model reduces the frequency of recommendation of items that are heavily rated by MC model (top-30 items) while those items that get no recommendation using MC algorithm also gain visibility via our design. Above results validate our models capability in improving the diversity of recommendations.

*4.5 Behavior of our Proposed Model-MCAD*

The convergence behavior of our optimization framework for one test instance is shown in fig 4. The smooth decrease in objective function value and the resultant convergence, together with convex nature of our formulation, validates our claim that a global minima is obtained. The same in turn ensures optimality of solution.

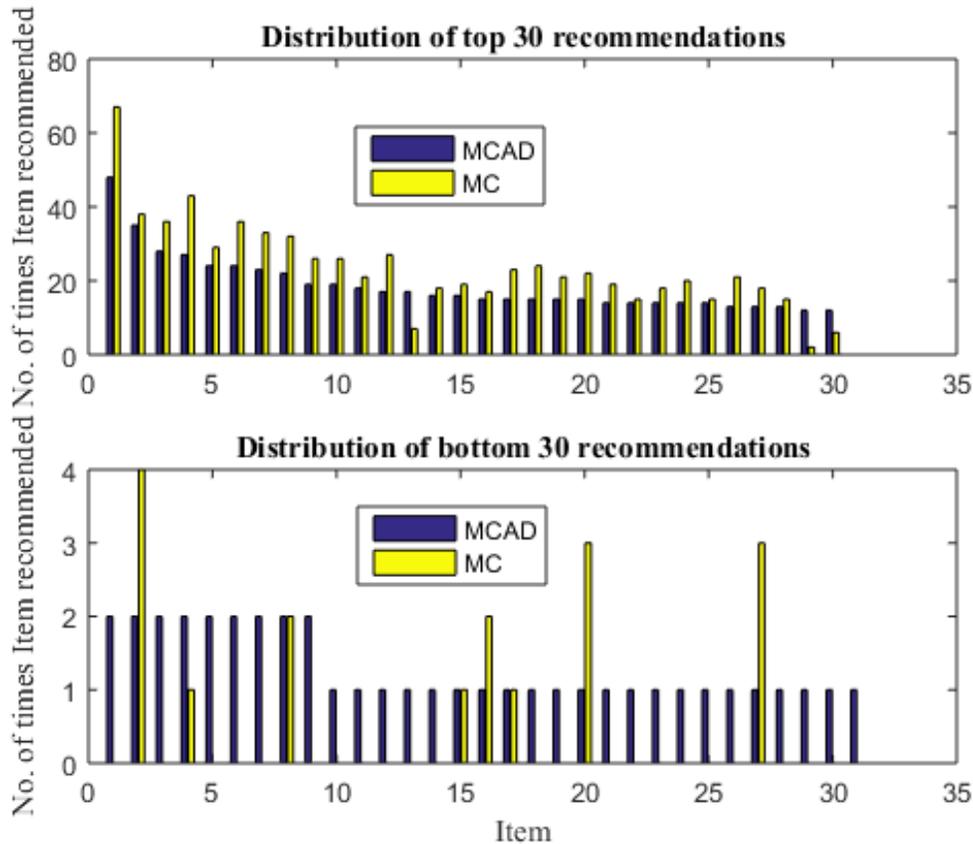

Fig 3. Distribution of Top-30 and bottom-30 recommended items

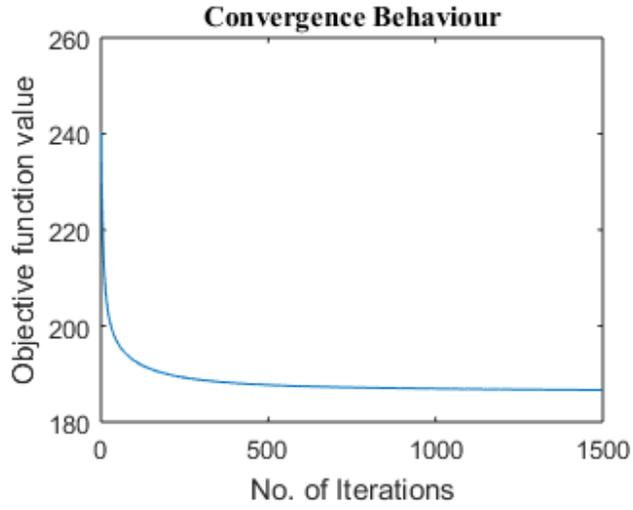

Fig. 4 Convergence Behavior of MCAD

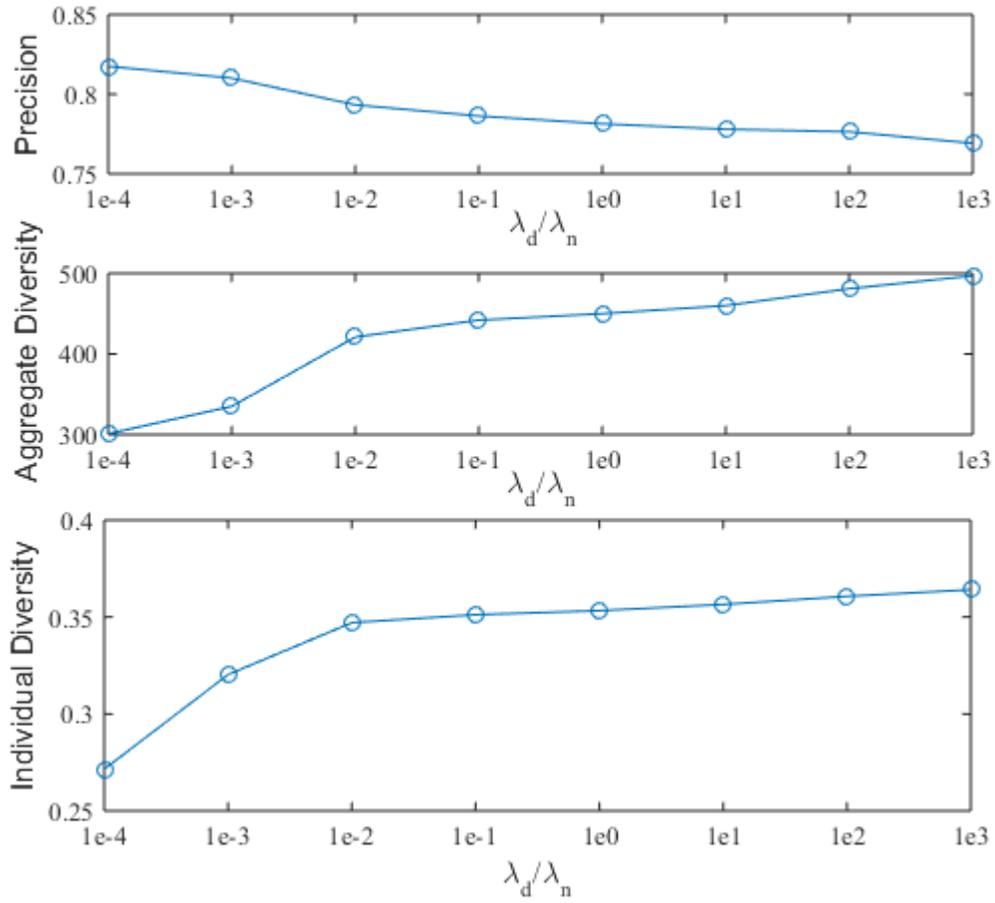

Fig. 5 Accuracy-Diversity Tradeoff

Fig. 5 shows the accuracy-diversity tradeoff achieved by varying the relative importance given to accuracy and diversity promoting constraints in our formulation (15) for the 100K Movielens dataset. The tradeoff between the two is a function of the relative values of the two regularization parameters $\lambda_n$ and $\lambda_d$. In Fig. 5 we show the change in precision, aggregate diversity and individual diversity as a function of the ration of two regularization parameters i.e. $\lambda_d \big/ \lambda_n$. It can be seen that as the relative importance of diversity promoting term increases, the precision drops while the diversity measures increase; in congruence with our design concept. It also shows the capability of our model to provide varying degree of diversity, with minimum loss of accuracy, as per the requirement of the RS under consideration.

The above two figures provide an empirical validation of our proposed model aimed at establishing an optimal accuracy-diversity tradeoff.

### 4.6 Comparison with Existing Techniques

In this section, we present the comparison of our approach with existing approaches aimed at improving diversity.

We compare the performance of our formulation against three existing schemes, Reverse predicted rating value (RPRV) [16], Item average rating (IA) [16] and clustering [23], all ranking strategies built on top of existing CF scheme (we used MC as the base CF technique because of its high precision).

All the (existing) techniques compared against use a modified ranking strategy (unlike highest rating being ranked first), wherein all items (movies) ranked above a predefined (heuristically selected) threshold ($T_{rank}$) are ranked as per the modified scheme. In RPRV, all items ranked above $T_{rank}$ are ranked in reverse order i.e. items with lower ratings ranked higher; to suggest less popular items. In IA scheme, items having lower average ratings (computed from known values) are ranked higher based on the argument that such items are less popular and hence recommending those increases diversity. In the work proposed in [23], hierarchical clustering was used to select items belonging to diverse genres.

Table 4 and 5 gives the percentage change in Gini coefficient (GC), novelty (NV), aggregate (AD) and individual diversity (ID) as a function of precision loss, obtained with all of the compared algorithms over the base matrix completion (MC) method. The results given are for a recommendation list of length 5. It can be observed from table 4 that for a given precision loss our formulation preforms significantly better than all other approaches compared against. For example, for a 5% reduction in precision, our algorithm yield a 76% jump in AD, 15% in ID and 33% in NV and a 7.5% reduction in GC . Corresponding values for next best algorithm are 61, 10, 16 and 1 percent. For 1M dataset (as seen from table 5), the superiority of our design is much more prominent. For example, for a 2% reduction in precision, our AD is 157% higher than base whereas RPRV gives a comparatively small increment of 65%. Similar trend is observed for individual diversity as well as novelty. The results for clustering based approach are really poor with precision sacrificed higher than gain in diversity.

TABLE 4. COMPARISON OF EXISTING ALGORITHMS WITH MCAD FOR 100K MOVIELENS DATASET (N=5)

| Precision Loss (% reduction) | MCAD (% change) | | | | MC-IA (% change) | | | | MC-RPRV (% change) | | | | MC-Clustering (% change) | | | |
|---|---|---|---|---|---|---|---|---|---|---|---|---|---|---|---|---|
| | Aggregate Diversity | Individual Diversity | Novelty | Gini Coeff. | Aggregate Diversity | Individual Diversity | Novelty | Gini Coeff. | Aggregate Diversity | Individual Diversity | Novelty | Gini Coeff. | Aggregate Diversity | Individual Diversity | Novelty | Gini Coeff. |
| 3 | 49.3 | 8.7 | 18.7 | 4.1 | 44.9 | 6.9 | 10.5 | 0.01 | 38.3 | 6.8 | 9.7 | 0.0 | 1.8 | 0.9 | 1.2 | 0.0 |
| 3.5 | 54.4 | 11.5 | 23.8 | 4.7 | 49.9 | 7.8 | 13.0 | 0.02 | 42.8 | 7.4 | 11.6 | 0.0 | 1.8 | 1.1 | 1.3 | 0.0 |
| 4 | 68.2 | 14.1 | 29.6 | 6.5 | 54.3 | 8.7 | 14.8 | 0.40 | 45.6 | 8.0 | 13.3 | 0.21 | 1.8 | 1.3 | 1.6 | 0.01 |
| 4.5 | 73.7 | 15.1 | 31.9 | 7.0 | 57.8 | 9.4 | 15.7 | 0.75 | 48.5 | 8.5 | 15.0 | 0.69 | 1.7 | 1.3 | 1.6 | 0.0 |
| 5 | 75.9 | 15.4 | 33.0 | 7.4 | 61.3 | 10.1 | 16.5 | 1.07 | 52.8 | 9.3 | 15.7 | 0.89 | 1.5 | 1.5 | 1.8 | 0.01 |
| 5.5 | 79.2 | 16.4 | 35.3 | 7.8 | 64.7 | 10.8 | 17.3 | 1.28 | 56.8 | 10.0 | 16.3 | 1.01 | 1.5 | 1.6 | 1.9 | 0.02 |
| 6 | 82.9 | 16.8 | 36.8 | 8.1 | 68.6 | 11.7 | 18.4 | 1.45 | 59.8 | 10.5 | 16.6 | 1.12 | 1.5 | 1.8 | 2.2 | 0.01 |
| 6.5 | 85.6 | 17.1 | 37.0 | 8.2 | 73.2 | 12.8 | 20.0 | 1.69 | 62.7 | 10.9 | 17.0 | 1.31 | 1.5 | 2.0 | 2.4 | 0.01 |
| Base MC | Precision: 0.8148, AD: 274, ID: 0.3152, Novelty: 1.3011, Gini Coeff: 0.9318 | | | | | | | | | | | | | | | |

TABLE 5. COMPARISON OF EXISTING ALGORITHMS WITH MCAD FOR 1M MOVIELENS DATASET (N=5)

| Precision Loss (% reduction) | MCAD (% change) | | | | MC-IA (% change) | | | | MC-RPRV (% change) | | | | MC-Clustering (% change) | | | |
|---|---|---|---|---|---|---|---|---|---|---|---|---|---|---|---|---|
| | Aggregate Diversity | Individual Diversity | Novelty | Gini Coeff. | Aggregate Diversity | Individual Diversity | Novelty | Gini Coeff. | Aggregate Diversity | Individual Diversity | Novelty | Gini Coeff. | Aggregate Diversity | Individual Diversity | Novelty | Gini Coeff. |
| 1 | 117.1 | 14.9 | 27.6 | 26.2 | 33.8 | 4.4 | 7.7 | 2.44 | 46.8 | 5.4 | 9.5 | 1.80 | 2.0 | 0.1 | 0.4 | 0.0 |
| 1.5 | 136.8 | 16.1 | 29.7 | 30.0 | 42.8 | 5.6 | 9.6 | 3.03 | 56.1 | 6.4 | 11.4 | 2.12 | 2.6 | 0.3 | 0.8 | 0.03 |
| 2 | 157.2 | 17.7 | 33.6 | 33.8 | 52.7 | 6.8 | 11.9 | 3.89 | 65.3 | 7.6 | 13.7 | 3.01 | 3.1 | 0.6 | 1.3 | 0.1 |
| 2.5 | 174.2 | 19.2 | 37.8 | 38.5 | 60.6 | 7.9 | 13.9 | 4.86 | 74.0 | 9.0 | 16.2 | 3.45 | 3.4 | 0.8 | 1.8 | 0.19 |
| 3 | 187.1 | 20.5 | 41.3 | 41.3 | 66.5 | 8.8 | 15.5 | 5.90 | 82.8 | 10.3 | 18.4 | 3.89 | 3.2 | 1.1 | 2.3 | 0.20 |
| 3.5 | 198.4 | 21.7 | 43.1 | 42.1 | 73.6 | 9.9 | 17.6 | 6.49 | 90.4 | 11.2 | 20.2 | 4.53 | 3.4 | 1.3 | 2.7 | 0.21 |

| Base MC | Precision: 0.8716,   AD: 653,   ID: 0.3459,   Novelty: 1.5584, Gini Coeff.: 0.9598 |
|---|---|

As stated previously, to showcase the effectiveness of our proposed formulation across varying number of recommendations, we briefly summarize the results for a recommendation list of length 10 and 20, i.e. N=10 and 20. We show the change in individual diversity and aggregate diversity for top-10 and top-20 recommendations as a function of drop in precision for 100K dataset in fig 6. We show the comparison with only IA and RPRV as they are the two better performing benchmarks. It can be observed that our model consistently works better than others across varying lengths of recommendation list. The percentage improvement in diversity as a function of precision for our model, compared to existing approaches (IA and RPRV), is consistent across varying number of recommendations (5 to 20). It can be seen that for all values of N considered (5, 10 and 20), the diversity obtained with our model is much higher compared

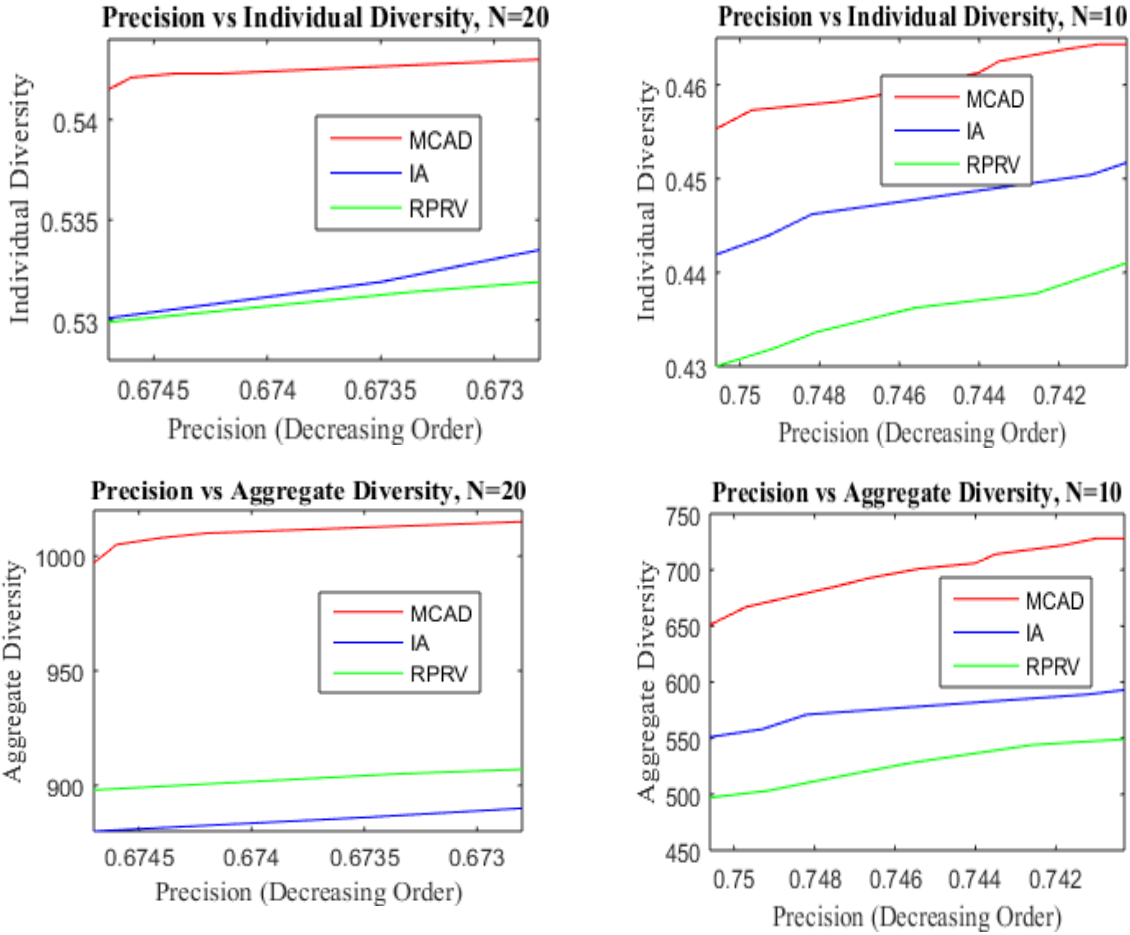

Fig 6. Variation in Individual and Aggregate diversity as a function of precision

to that achieved with the other two methods for the same accuracy levels.

Thus, it is evident that our formulation consistently performs significantly better than other state of the art approaches. Similar trend is observed for 1M datasets, however the results for the same are not shown here.

## 5  CONCLUSION

In this work we propose a novel formulation for achieving accuracy-diversity balance. A certain level of diversity with decently high precision improves user satisfaction as well as profits for business portals.

Earlier works in this direction proposed heuristic ranking strategies implemented on top of existing collaborative filtering schemes. In this work, we propose a combined formulation which promotes increase in diversity with a very small reduction in precision. Our proposed model is a convex framework built as a modification of matrix completion formulation. All the existing works either use less efficient neighbourhood based methods or non-convex matrix factorization framework.

Our novel formulation is based on the notion that for enhanced diversity, each item category should have equal probability of being recommended. We incorporate this idea as a suitable regularization term to base matrix completion formulation; which inherently im-proves prediction accuracy. Also, unlike most of existing works, our proposed method simultaneously im-proves both individual and aggregate diversity.

Experimental evaluation on the real world movie database corroborates our claim that our design improves significantly upon the existing state of the art approaches. Also, our design does away with the need for selecting ranking thresholds arbitrarily and choice of heuristic ranking strategies. Being a convex formulation, it ensures attainment of optimum solution unlike existing approaches.